\documentstyle[twocolumn,prl,aps,floats]{revtex}
\begin{document}
\input{psfig.sty}
\draft

\title{Free-space quantum key distribution}

\author{W. T. Buttler,$^{1}$ R. J. Hughes,$^{1}$ P. G. Kwiat,$^{1}$ 
G. G. Luther,$^{1}$ G. L. Morgan,\\$^{1}$ J. E. Nordholt,$^{2}$ 
C. G. Peterson,$^{1}$ and C. M. Simmons$^{1}$}

\address{University of California,\\Los Alamos National Laboratory, 
\\$^{1}$Physics Division,\\$^{2}$Nonproliferation and International 
Security,\\Los Alamos, NM 87545}

\date{\today}
\maketitle

\begin{abstract}
A working free-space quantum key distribution (QKD) system has been developed 
and tested over a 205-m indoor optical path at Los Alamos National Laboratory
under fluorescent lighting conditions. Results show that free-space QKD can 
provide secure real-time key distribution between parties who have 
a need to communicate secretly.
\end{abstract}


Quantum cryptography was introduced in the mid-1980s \cite{BB84} as a new 
method for generating the shared, secret random number sequences, or 
cryptographic keys, that are used in crypto-systems to provide communications 
security. The appeal of quantum cryptography is that its security is based on 
laws of Nature, in contrast to existing methods of key distribution that 
derive their security from the perceived intractability of certain problems 
in number theory, or from the physical security of the distribution process. 
Since the introduction of quantum cryptography, several groups have 
demonstrated that quantum key distribution (QKD) can be performed over 
multi-kilometer distances of optical fiber \cite{1_km}-\cite{Aero97}, but 
the utility of the method would be greatly enhanced if it could also be 
performed over free-space paths, such as are used in laser communications 
systems. Indeed there are certain key distribution problems in this category 
for which QKD would have definite practical advantages (for example, it is 
impractical to send a courier to a satellite). We are developing QKD for use 
over line-of-sight paths, including surface to satellite, and here we report 
our first results on key generation over indoor paths of up to $205$ m. 

The feasibility of QKD over free-space paths might be considered problematic 
because it requires the transmission of single photons through a medium with 
varying properties and detection of these photons against a high background. 
However, others have shown that the combination of sub-nanosecond timing, 
narrow filters \cite{PhotonByPhoton,DaylightPairs}, and spatial filtering can 
render both of these problems tractable. Furthermore, the atmosphere is 
essentially non-birefringent at optical wavelengths, allowing faithful
transmission of the single-photon polarization states used in QKD.

A QKD procedure starts with the sender, ``Alice,'' generating a secret random 
binary number sequence. For each bit in the sequence, Alice prepares and 
transmits a single photon to the recipient, ``Bob,'' who measures each 
arriving photon and attempts to identify the bit value Alice has transmitted. 
Alice's photon state preparations and Bob's measurements are chosen from sets 
of non-orthogonal possibilities. For example, in the B92 protocol \cite{B92} 
Alice agrees with Bob (through public discussion) that she will transmit a 
horizontally-polarized photon, $|h\rangle$, for each ``0'' in her 
sequence, and a right-circular-polarized photon, $|rcp\rangle$, for 
each ``1'' in her sequence. Bob agrees with Alice to randomly test the 
polarization of each arriving photon in one of two ways: he either tests 
with vertical polarization, $|v\rangle$, to reveal ``1s,'' or left-circular 
polarization, $|lcp\rangle$, to reveal ``0s.'' Note that Bob will never 
detect a photon for which he and Alice have used a preparation/measurement 
pair that corresponds to different bit values, such as $|h\rangle$ and 
$|v\rangle$, which happens for $50$\% of the bits in Alice's sequence. However, 
for the other $50$\% of Alice's bits where the preparation and measurement 
protocols agree, such as $|h\rangle$ and $|lcp\rangle$, there is a 50\% 
probability that Bob detects the photon, as shown in TABLE I. So, by 
detecting photons Bob is able to identify a random $25$\% portion of 
the bits in Alice's sequence, assuming no bit loss in transmission or 
detection. (This $25$\% efficiency factor is the price that Alice and Bob 
must pay for secrecy.) Bob then communicates to Alice over a public channel 
the locations, but not the bit values, in the sequence where he detected 
photons, and Alice retains only these detected bits from her initial 
sequence. The resulting detected bit sequences are the raw key material 
from which a pure key is distilled using classical error detection 
techniques. An eavesdropper, ``Eve,'' can neither ``tap'' the key 
transmissions, owing to the indivisibility of a photon 
\cite{indivis1,indivis2}, nor copy them owing to the quantum ``no-cloning'' 
\cite{noclone1}-\cite{noclone4} theorem. Furthermore, the non-orthogonal 
nature of the quantum states ensures that if Eve makes her own measurements 
she will be detected through the elevated error rate she causes by 
the irreversible ``collapse of the wavefunction'' \cite{Eavesdropping}.

The prototype QKD transmitter (FIG. 1) consisted of a temperature controlled 
diode laser, a collimating lens, two dielectric mirrors, a fiber to 
free-space launch system, a single-mode fiber pigtailed polarization 
neutral beamsplitter, a variable optical attenuator (OA), a $\sim10$-m 
single-mode optical fiber delay, a $2.5$ nm bandwidth interference filter 
(IF), a polarizing beamsplitter (PBS), a low-voltage pockels cell (PC), 
and an $8\times$ beam expander (BE).

The diode laser wavelength is temperature selected to $772$ nm, and the laser 
is configured to emit a short, weak coherent pulse of $\sim1$ ns 
length, containing approximately $10^{5}$ photons.

The free-space QKD receiver (FIG. 2) was comprised of a $3.5$ in. Cassegrain 
telescope (CT), a free-space to fiber launch system, a single-mode fiber 
pigtailed polarization neutral beamsplitter, two sets of polarization 
controllers (each consisting of a quarter-wave retarder and a half-wave 
retarder), a PBS, and a single photon counting module, or SPCM (EG\&G part 
number: SPCM-AQ 142-FL). The prototype receiver did not include an 
interference filter but it is expected that future versions of the receiver 
will incorporate this feature to reduce background light levels.

A computer control system, ``Alice,'' starts the QKD protocol by pulsing 
the diode laser at a rate previously agreed upon between herself and the 
receiving computer control system, ``Bob.'' Each laser pulse is launched 
into a single-mode optical fiber and then split by the beamsplitter with 
equal probability between the direct path and the delay path. The direct 
path produces a coherent ``bright pulse'' of $\sim10^{5}$ photons which 
Bob uses as his system trigger for timing purposes.

Light traveling along the direct path passes through the IF, the PBS, the PC, 
and is then launched into free-space from the BE. The IF constrains 
wavelength, and the PBS is oriented to transmit horizontal polarization.

The fiber delay and OA are used to delay the diverted pulse by $\sim50$ ns 
as well as attenuate the diverted pulse to an average of $\sim1.4$ photons 
per pulse. This attenuated pulse then impinges again upon the beamsplitter, 
which transmits a dim-pulse with and average of $\sim0.7$ photons that 
follows the bright pulse along the direct path through the IF, the PBS, 
the PC, and the BE. (The attenuated pulse only approximates a 
``single-photon'' state; we tested out the system with an average of 
$\sim0.7$ photons per ``dim-pulse.'' This corresponds to a $2$-photon 
probability of $\sim12$\% and implies that $\sim30$\% of the detectable 
dim-pulses will contain $2$ or more photons, e.g., with a Poisson 
distribution with an average photon number of $0.7$ there will be $\sim50$ 
empty sets, $\sim35$ sets of $1$ photon, $\sim12$ sets of $2$ photons, 
and $\sim3$ sets of $3$ photons for every $100$ dim-pulses.) The PBS 
transmits the $|h\rangle$ dim-pulse to the PC which is randomly switched 
to affect only the dim-pulse polarization. The random switch setting is 
determined by discriminating a random voltage generated by a white noise 
source and either passes the dim-pulse unchanged as $|h\rangle$ (zero-wave 
retardation) or changes it to $|rcp\rangle$ (quarter-wave retardation), 
depending on the random bit value. The bright pulse's polarization is 
never altered.

Bob then collects the bright- and dim-pulses with the Cassegrain telescope 
and launches them into single-mode fiber. The bright pulse is split at the 
beamsplitter along two independent paths---one path [the long path (LP)] 
is approximately $5$ ns longer than the other path [the short path (SP)]. 
Each path contains polarization controlling optics which terminate upon 
the PBS. We configured our system to operate with a single SPCM, but we 
have also operated with SPCMs at both of the output ports of the PBS.

If the dim-pulse of $\sim0.7$ photons is collected and launched into the 
fiber at the receiver it will be diverted by the beamsplitter with equal 
probability along one of the two possible optical paths. In the prototype 
system the polarization controlling optics were adjusted to behave together 
as a quarter-wave retarder along the SP, and a zero-wave retarder along the 
LP. Thus, a dim-pulse of $|rcp\rangle$ traveling the SP is converted to 
$|v\rangle$ and reflected away from the SPCM. Conversely, a 
dim-pulse of $|h\rangle$ traveling the SP is converted to $|rcp\rangle$ 
and is transmitted toward or reflected away from the SPCM with equal 
probability. Similarly, a dim-pulse of $|h\rangle$ traveling the LP is 
transmitted away from the SPCM, but a dim-pulse of $|rcp\rangle$ 
is reflected toward or transmitted away from the SPCM with equal probability.

We used the differing path lengths, together with fast timing electronics 
gated with narrow coincidence windows ($\sim5$ ns), to determine dim-pulse 
polarizations with a single detector. Specifically, a coincidence observed 
$50$ ns after the bright pulse (early coincidence) informs Bob that the 
dim-pulse was of $|rcp\rangle$, while a coincidence observed $55$ ns after 
the bright pulse (late coincidence) tells Bob that the dim-pulse was of 
$|h\rangle$. The detector dead time was $\sim35$ ns.

A variety of transmitter and receiver configurations were used to evaluate 
the equipment and test out the optical elements over optical path lengths 
of $2$-, $36$-, and $205$-m, but here we discuss only the $205$-m results. The 
$205$-m experiment was performed with the transmitter and receiver colocated 
in order to simplify data acquisition. The $205$-m optical path was achieved 
by sending the emitted beam up and down a $\sim17.1$-m laboratory hallway $6$ 
times with the use of $10$ mirrors, and a corner cube under fluorescent 
lighting conditions.

The corner cube was used to determine the feasibility of transmitting 
single photons from a ground station to a low earth orbit satellite covered 
with corner cubes (such as LAGEOS-I and LAGEOS-II) and back. (Note: the 
primary property of the corner cube is its ability to return light back 
along the path it came. However, the corner cube also possesses the 
seldom-noted feature that each of its $6$ possible optical paths will 
transform a given incident polarization differently\cite{corner_cube}. 
Because of this, a fully illuminated corner cube cannot be used to perform 
polarization dependent experiments. Therefore, only one path through the 
corner cube was used during the experiment.)

The coupling efficiency, $\eta$, between the transmitter and receiver for 
the $205$-m path was $\eta\sim2$\%, where $\eta$ accounts for losses between 
the transmitter and the power coupled into the single-mode fibers preceding 
the detector at the receiver. This efficiency led to a bit-rate of 
$\sim50$ Hz when the transmitter was pulsed at a rate of $\sim20$ kHz over 
the $205$-m path, with the system operating at an average of $\sim0.7$ 
photons per dim-pulse. The final bit-rate is the product of $\eta$ 
and the probabilities that the weak coherent pulse of photons will reach 
the detector, and the probabilities that the detector will fire when 
Poisson distributed photons reach the detector. [The detector efficiency is 
a function of the average photon number per dim-pulse, and accounts 
for the probability the detector will fire given 1 photon ($p(1)\sim0.65$), 
$2$ photons ($p(2)\sim0.878$), $3$ photons ($p(3)\sim0.957$), etc., reach the 
detector. These probabilities are convolved with the probabilities that 
$1$, $2$, $3$, or more of those photons actually reach the detector and then 
convolved with the Poisson probabilities for $1$, $2$, $3$, or more photons 
per dim-pulse ($p(1)=0.348$, $p(2)\sim0.123$, $p(3)\sim0.0284$, etc.). These 
convolutions are then summed to give the detection efficiency as a function 
of the Poisson average photon number.] The bit error rate (BER) for the 
$205$-m path was $\sim6$\%, where the BER is defined as the ratio of the 
bits received in error to the total number of bits received. A sample of raw 
key material from the $205$-m experiment, with errors, is shown in TABLE II.

The narrow coincidence time windows in Bob's receiver minimized bit errors due 
to detector dark noise ($\sim80$ Hz); the ambient background was $\sim1$ kHz. 
These low noise rates amounted to $\sim1$ bit-error every $9$ s. After-pulsing 
of the SPCMs caused by the bright pulses contributed $\sim2$\% to the 
total BER---an average rate of $1$ bit-error per second. In addition, bright 
pulse reflections within the transmitter caused the ``1s'' errors (late 
coincidence errors) to be about $6$ times higher than the ``0s'' errors 
(early coincidence). After-pulsing errors could be reduced by increasing 
the length of the fiber delay to further separate the bright and dim 
pulse and should result in a BER of $\sim4$\%---an average rate of $2$ 
bit-errors per second; reflection errors could be reduced through the use of 
angle polished fiber termination and should result in a BER of $\sim2$\%. It 
is important to eliminate reflection errors because these are weaknesses which 
could be exploited by Eve. The BER might be further reduced to $\sim1$\% 
by elimination of the common PBS at the receiver, and by operating 
the receiver in a $2$ detector configuration. The poor coupling efficiency 
($\eta\sim2$\%) together with the constant average bit-errors caused by 
after-pulsing and reflections (about $3$ bit-errors per second) prevented 
us from effectively operating the prototype system below an average of 
$\sim0.7$ photons per dim-pulse.

This experiment implemented a two-dimensional parity check scheme which 
allowed the generation of error free key material. The error detection 
program permitted the isolation of error free bits from key material with 
BERs exceeding $10$\%. A further stage of ``privacy amplification'' is 
necessary to reduce any partial knowledge gained by an eavesdropper to 
less than $1$-bit of information \cite{PrivAmp}. We have not implemented 
this protocol at this time. Our prototype incorporates a ``one time 
pad\footnote{One time pad encryption utilizes a unique random string of 
key bits to encrypt a single plaintext message. In particular, the key 
bit string exactly the same length as the plaintext string and is used 
only one time. Encryption (decryption) is accomplished by XORing the 
message bits (encrypted bits) with the key bits.}'' \cite{Vernam} 
encryption (also known as the Vernam Cipher)---the only provably secure 
encryption method, and could also support any other symmetric key system.

The original form of the B92 protocol \cite{B92} has a weakness to a 
``man-in-the middle,'' or opaque, attack by Eve. For instance, Eve could 
measure Alice's photons in Bob's basis and only send a photon, or coherent 
photon pulse, when she identifies a bit. However, if Eve retransmits each 
observed bit as a single-photon (or a weak coherent pulse) she will 
noticeably lower Bob's bit-rate. To compensate for the additional 
attenuation to Bob's bit-rate Eve could send on a coherent photon pulse 
of an intensity appropriate to raise Bob's bit-rate to a level similar 
to her own bit-rate with Alice. [In fact, if Eve sends a bright classical 
pulse (a pulse of a large average photon number) she guarantees that 
Bob's bit-rate is equal to her own.] Our system protects against this 
scenario when operated with two SPCMs. For example, this type of attack would 
be revealed by an increase in ``dual-fire'' errors which occur when both 
SPCMs fire simultaneously. (In a perfect system there would be no 
``dual-fire'' errors, regardless of the average photon number per pulse, but 
in an imperfect experimental system, where bit-errors occur, dual-fire errors 
will occur.) A better protection would be to use the BB84 \cite{BB84} 
protocol, which our system also supports.

Over the next few months we intend to implement design changes to the 
transmitter and receiver in order to increase system efficiency, $\eta$, 
and increase the total range of the QKD system. Our calculations show that 
a narrow filter, the spatial filtering and the narrow coincidence timing 
provided by this system will allow reliable key distribution under bright 
daylight conditions. Our goal is to exchange key bits outdoors over one or 
two kilometers by the end of $1997$.

Finally, we note that somewhat similar results to these presented here are 
reported in reference \cite{FreeSpace}. However, the protocol of reference 
\cite{FreeSpace} was implemented with a modulated HeNe laser, utilized long 
pulse lengths ($\sim100$ ns), and active polarization switching at the 
receiver, whereas we implemented our protocol over a line-of-sight path 
$35$\% longer than in reference \cite{FreeSpace} with a system which 
incorporates short pulse lengths ($\sim1$ ns, and allows the use of narrow 
coincidence timing windows to minimize ambient background noise allowing 
daytime applications) and passive polarization switching at the receiver 
(a simpler design than in reference \cite{FreeSpace} which will be critical 
for the locating of a receiver, Bob, on a satellite).

R. J. H. wishes to extend special thanks to J. G. Rarity for the many 
helpful discussions regarding free-space quantum cryptography, and W. T. B. 
extends his appreciation to S. K. Lamoreaux and A. G. White for their 
helpful conversations. The work described in this letter was performed 
with U. S. Government funding.
%\bigskip

\newpage
%\begin{table}[!ht]
\begin{table}
\caption{Observation Probabilities}
\begin{center}
\begin{tabular}{|l|c|c|c|c|} \hline
Alice's Bit Value & ``0'' & ``0'' & ``1'' & ``1'' \\
Bob Tests With    & ``1'' & ``0'' & ``1'' & ``0'' \\ \hline
Observation Probability & p$=0$ & p$=\frac{1}{2}$ & p$=\frac{1}{2}$ & p$=0$ \\ 
\hline
\end{tabular}
\end{center}
\end{table} 

%\begin{table}[!ht]
\begin{table}
\caption{A 200-Bit Sample of Alice's (A) and Bob's (B) Raw Key Material 
Generated by Free-Space QKD}
\begin{center}
\begin{tabular}{|l|l|l|l|l|l|} \hline
A & 11111010 & 10100100 & 00110010 & 10011011 & 01010110 \\
B & 1111101{\bf 1} & 1{\bf 1}100{\bf 0}00 & 0011{\bf 1}010 & 10011011 & 
01010110 \\ \hline
A & 00111101 & 01101111 & 11010000 & 01101111 & 01011011 \\
B & 00111101 & 01101111 & 11010000 & 01101111 & 01011011 \\ \hline
A & 11100100 & 01010001 & 10110100 & 10110101 & 01101011 \\
B & 11100100 & 01010001 & {\bf 0}0{\bf 0}10100 & 10110101 & 01101011 \\ \hline
A & 10001011 & 11010111 & 10101110 & 10100111 & 00010011 \\
B & {\bf 0}00010{\bf 0}1 & 11010111 & 10101{\bf 0}10 & 10100111 & 00010011 \\ 
\hline
A & 01000010 & 00100011 & 00111001 & 01101100 & 01110001 \\
B & 01000010 & 00100011 & 00111001 & 01101{\bf 0}00 & 01110001 \\ \hline
\end{tabular}
\end{center}
\end{table}

\begin{figure}
\psfig{figure=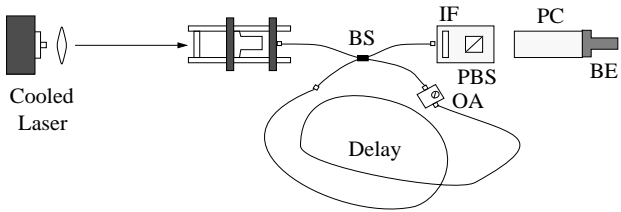,width=3.375 in}
\caption{Free-Space QKD Transmitter (Alice).}
\label{app}
\end{figure} 

\begin{figure}
\psfig{figure=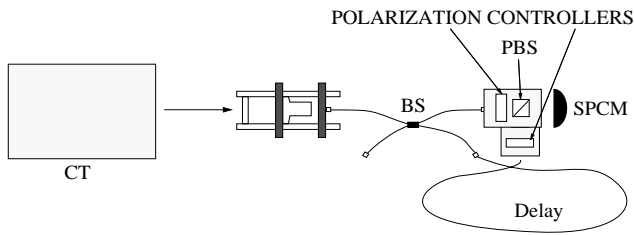,width=3.375 in}
\caption{Free-Space QKD Receiver (Bob).}
\label{a1}
\end{figure}


\begin{thebibliography}{99}
\bibitem{BB84} Bennett, C. H., and G. Brassard, Proc. of IEEE Int. Conf. on 
Comp., Sys., and Sig. Proc., Bangalore, India, 175 (1984).

\bibitem{1_km} Muller, A., J. Breguet, and N. Gisin, Europhys. Lett., 
{\bf 23}, 383-388 (1993).

\bibitem{TRT} Townsend, P. D., J. G. Rarity, and P. R. Tapster, Elec. 
Lett., {\bf 29}, 634-635 (1993).

\bibitem{Franson} Franson, J. D., and H. Ilves, Appl. Opt., {\bf 33}, 
2949-2954 (1994).

\bibitem{ContempPhys} Hughes, R. J., D. M. Alde, P. Dyer, G. G. Luther, 
G. L. Morgan, and M. Schauer, Contemp. Phys., {\bf 36}, 149-163 (1995).

\bibitem{Marand} Marand, C., and P. D. Townsend, Opt. Lett., {\bf 20}, 
1695-1697 (1995).

\bibitem{23_km} Muller, A., H. Zbinden, and N. Gisin, Europhys. Lett., 
{\bf 33}, 335-339 (1996).

\bibitem{LectureNotes} Hughes, R. J., G. G. Luther, G. L. Morgan, C. G. 
Peterson, and C. Simmons, Lecture Notes In Computer Science, {\bf 1109}, 
329-338 (1996).

\bibitem{Aero97} Hughes, R. J., W. T. Buttler, P. G. Kwiat, G. G. Luther, 
G. L. Morgan, J. E. Nordholt, C. G. Peterson, and C. M. Simmons, Proc. of 
SPIE, {\bf 3076}, 2-11 (1997).

\bibitem{PhotonByPhoton} Walker, J. G., S. F. Seward, J. G. Rarity, and 
P. R. Tapster, Quant. Opt. {\bf 1}, 75-82 (1989).

\bibitem{DaylightPairs} Seward, S. F., P. R. Tapster, J. G. Walker, and 
J. G. Rarity, Quant. Opt., {\bf 3}, 201-207 (1991).

\bibitem{B92} Bennett, C. H., Phys. Rev. Lett. {\bf 68}, 3121-3124 (1992).

\bibitem{indivis1} Clauser, J. F., Phys. Rev. D, {\bf 9}, 853-860 (1974). 

\bibitem{indivis2} Grangier, P, G. Roger, A. Aspect, Europhys. Lett. 
{\bf 1}, 173-179 (1986). 

\bibitem{noclone1} Wooters, W. K., and W. H. Zurek, Nature {\bf 299}, 
802-803 (1982). 

\bibitem{noclone2} Dieks, D., Phys. Lett. A, {\bf 92}, 271-272 (1982).

\bibitem{noclone3} Milonni, P. W., and M. L. Hardies, Phys. Lett. A, 
{\bf 92}, 321-322 (1982). 

\bibitem{noclone4} Mandel, L., Nature {\bf 304}, 188-188 (1983).

\bibitem{Eavesdropping} Ekert, A. K., B. Huttner, G. M. Palma, and A. 
Peres, Phys. Rev. A, {\bf 50}, 1047-1056 (1994).

\bibitem{corner_cube} Liu, J, and R. M. Azzam, App. Opt., {\bf 36} 
1553-1559 (1997).

\bibitem{PrivAmp} Bennett, C. H., G. Brassard, C. Crepeau. and U. M. 
Maurer, IEEE Trans. Inf. Th., {\bf 41}, 1915-1923 (1995).

\bibitem{Vernam} Vernam, G. S., Trans. Am. Inst. Elect. Eng., {\bf XLV}, 
295-301 (1926).

\bibitem{FreeSpace} Jacobs, B. C., and J. D. Franson, Opt. Lett., 
{\bf 21}, 1854-1856 (1996).

\end{thebibliography}
\end{document}